\documentclass[12pt]{article}
\usepackage{amsfonts}
\usepackage[T1]{fontenc}
\usepackage[utf8]{inputenc}
\usepackage[english]{babel}
\usepackage{amsmath}
\usepackage{theorem}
\usepackage{enumerate}
\usepackage{pxfonts}
\usepackage{txfonts}
\DeclareMathSizes{14}{14}{14}{14}

\newtheorem{theorem}{Theorem}

\newtheorem{lemma}{Lemma}

\newtheorem{defi}{Definition}
\newtheorem{remark}{Remark}
\usepackage[unicode,colorlinks,plainpages=false]{hyperref}
\begin{document}

\centerline{\bf Boolean-type Retractable State-finite Automata Without Outputs}
\medskip
\centerline{ M\'ark F\"uzesdi}
\medskip
\centerline{Department of Algebra}
\centerline{Budapest University of Technology and Economics}
\centerline{e-mail: fuzesdim@math.bme.hu}
\bigskip

\begin{abstract}
 An automaton $\bf A$ is called a retractable automaton if, for every subautomaton $\bf B$ of $\bf A$, there is at least one homomorphism of $\bf A$ onto $\bf B$ which leaves the elements of $B$ fixed (such homomorphism is called a retract homomorphism  of $\bf A$ onto $\bf B$). We say that a retractable automaton \textbf{A}=(A,X,$\delta$) is Boolean-type if there exists a family $\{\lambda_B \mid \textrm{\textbf{B} is a subautomaton of \textbf{A}}  \}$ of retract homomorphisms $\lambda _B$ of $\bf A$ such that, for arbitrary subautomata $\textbf{B}_1$ and $\textbf{B}_2$ of $\bf A$, the condition $B_1\subseteq B_2$ implies $Ker\lambda _{B_2}\subseteq Ker\lambda _{B_1}$. In this paper we describe the Boolean-type retractable state-finite automata without outputs.
\end{abstract}

\medskip

\begin{section}{Introduction and motivation}
Let ${\bf A}=(A, X, \delta)$ be an automaton without outputs.
A subautomaton $\bf B$ of $\bf A$ is called a retract subautomaton if there is a homomorphism of $\bf A$ onto $\bf B$ which leaves the elements of $B$ fixed. A homomorphism with this property is called a retract homomorphism of $\bf A$ onto $\bf B$.

In \cite{nagy3}, A. Nagy introduced the notion of the retractable automaton. An automaton $\bf A$ (without outputs) is called a retractable automaton if every subautomaton of $\bf A$ is a retract subautomaton. He proved (in Theorem 3 of \cite{nagy3}) that if the lattice ${\cal L}({\bf A})$ of all congruences of an automaton $\bf A$ is complemented then $\bf A$ is a retractable automaton. He also defined the notion of the Boolean-type retractable automaton. We say that a retractable automaton \textbf{A}=(A,X,$\delta$) is Boolean-type if there exists a family $\{\lambda_B \mid \textrm{\textbf{B} is a subautomaton of \textbf{A}}  \}$ of retract homomorphisms $\lambda _B$ of $\bf A$ such that, for arbitrary subautomata $\textbf{B}_1$ and $\textbf{B}_2$ of $\bf A$, the condition $B_1\subseteq  B_2$ implies $Ker \lambda _{B_2}\subseteq Ker \lambda _{B_1}$. He proved (in Theorem 5 of \cite{nagy3}) that if the lattice ${\cal L}({\bf A})$ of all congruences of an automaton $\bf A$ is a Boolean algebra then $\bf A$ is a Boolean-type retractable automaton.

In \cite{nagy3}, A. Nagy investigated the not necessarily state-finite Boolean-type retractable automata containing traps (a state $c$ is called a trap of an automaton \textbf{A}=(A,X,$\delta$) if $\delta(c,x)=c$ for every $x \in X$). He proved that every Boolean-type retractable automaton
containing traps has a homomorphic image which is a Boolean-type retractable automaton containing exactly one trap. Moreover, he gave a complete description of Boolean-type retractable automata containing exactly one trap.

In \cite{nagy2}, the authors defined the notion of the strongly retract extension of automata. They proved that every state finite Boolean-type retractable automaton without outputs is a direct sum of Boolean-type retractable automata whose principal factors form a tree. Moreover, a state-finite automaton $\bf A$ is a Boolean-type retractable automaton whose principal factors form a tree if and only if it is a strongly retract extension of a strongly connected subautomaton of $\bf A$ by a Boolean-type retractable automaton containing exactly one trap (which is described in \cite{nagy3}).

In \cite{nagy3} and \cite{nagy2}, some theorem gives only necessary conditions for special retractable or Boolean-type retractable state-finite automata without outputs. Paper \cite{nagy1} is the first one which gives a complete description of state-finite retractable automata without outputs. Using the results of \cite{nagy1}, we give a complete description of Boolean-type retractable state-finite automata without outputs.
\end{section}

\begin{section}{Basic notations}

 By an automaton without outputs we mean a system $(A,X, \delta)$ where A and X are non-empty sets, and $\delta$ maps from the Cartesian product $A \times X$ to $A$. We will refer to A, X and $\delta$ as the state set, the input set and the transition function of $\bf{A}$, respectively. An automaton \textbf{A} is said to be state-finite, if the set A is finite. In this paper  by an automaton we always mean a state-finite automaton without outputs. We will follow the definitions and notations of \cite{nagy1}.
\medskip
\\ An automaton \textbf{B}=(B,X,$\delta_B$) is called a subautomaton of an automaton $\textbf{A}=(A, X, \delta)$ if B is a subset of A and $\delta_B$ is the restriction of $\delta$ to $B \times X$. A subautomaton $\bf B$ of an automaton $\bf A$ contained by every subautomaton of $\bf A$ is called the kernel of $\bf A$.

\medskip
By a homomorphism of an automaton $(A, X, \delta)$ into an automaton $(B, X, \gamma)$ we mean a map $\phi$ of the set $A$ into the set $B$ such that $\phi(\delta(a,x))=\gamma(\phi(a),x)$ for all $a\in A$ and $x\in X$.

\medskip

A congruence of an automaton $(A, X, \delta)$ is an equivalence $\alpha$ of the set $A$ such that, for all $a, b\in A$ and $x\in X$, the assumption $(a, b)\in\alpha$ implies $(\delta(a,x),\delta(b,x))\in ~ \alpha $. A congruence class $\alpha$ containing $a \in A$ will be denoted by $[a]_\alpha$.
The kernel of a homomorphism $\phi : (A,X,\delta)\mapsto (B, X, \gamma)$, which is denoted by $Ker\phi$, is defined as the following relation of $A$:
$Ker\phi \coloneqq \{(a,b)\in A\times A:\ \phi(a)=\phi(b)\}$. It is clear that $Ker\phi$ is a congruence on A.
\medskip
\\ We will denote the lattice of all congruences of an automaton $\bf A$ by ${\cal L}(A)$. For every $\alpha ,\beta \in {\cal L}(A)$,
$\alpha~\wedge~\beta~\coloneqq ~\alpha ~\cap~ \beta$ and $\alpha \vee \beta = (\alpha \cup \beta)^T $ where
\[(\alpha \cup \beta)^T=(\alpha \cup \beta)\cup ((\alpha \cup \beta)\circ (\alpha \cup \beta))\cup \dots \] is the transitive closure of $\alpha \cup \beta$ (here $\circ$ denotes the usual operation on the semigroup of all binary relations on $A$ (see \cite{clifford}) ).

\medskip
Let \textbf{B}=(B,X,$\delta_B$) be a subautomaton of an automaton \textbf{A}=$(A,X,\delta)$. The relation \newline $\varrho _B=\{ (b_1,b_2)\in ~ A\times ~ A: b_1=b_2 \quad \hbox{or}\quad b_1, b_2\in B\}$ is a congruence on \textbf{A}. This congruence is called the Rees congruence on $\bf A$ defined by $\bf B$. The $\varrho _B$-classes of $A$ are $B$ itself and every one-element set $\{a\}$ with $a\in A\setminus B$.
\end{section}

\begin{section}{Retractable automata}

\begin{defi} A subautomaton \textbf{B} of an automaton  \textbf{A}=$(A,X,\delta)$ is called a retract subautomaton if there exist a homomorphism $\lambda_B$ of $\bf A$ onto $\bf B$ which leaves the elements of $B$ fixed. An automaton is called retractable if its every subautomaton is retract. \cite{nagy3} \end{defi}

\begin{theorem}
A Rees-congruence $\varrho_B$ defined by a subautomaton \textbf{B}=$(B,X,\delta_B)$ of an automaton \textbf{A}=$(A,X,\delta)$ has a complement in the lattice $({\cal L}(A), \vee, \wedge)$ if and only if \textbf{B} is a retract subautomaton.
\end{theorem}
\medskip
\noindent \textbf{Proof} Let \textbf{A}=$(A,X,\delta)$ be an automaton. Assume that $\bf B$ is a subautomaton of $\bf A$ such that the Rees congruence $\varrho _B$ has a complement in ${\cal L}(A)$. By the proof of Theorem 3 of \cite{nagy3}, $\bf B$ is a retract subautomaton of $\bf A$.
Conversely, assume that $\bf B$ is a retract subautomaton of $\bf A$. We will show that the kernel of a retract homomorphism of $\bf A$ onto $\bf B$ is the a complement of the Rees congruence $\varrho_B$ defined by $\bf B$. We show this by proving that, for every states $a\neq b$ of $\bf A$, we have $(a,b)\notin \eta_B \wedge \varrho_B $ and $(a,b)\in \eta_B \vee
\varrho_B$ (here $\lambda_B$ denotes the corresponding retract homomorphism of \textbf{A} onto \textbf{B} and $\eta_B
\coloneqq Ker\lambda_B$). Let $a, b$ be arbitrary elements in $A$ with the condition $a\neq b$.
\begin{itemize} \item Case $a, b\in B$. \end{itemize}
Then $(a, b)\notin \eta_B \Rightarrow (a, b)\notin \eta_B \cap \varrho_B = \eta_B \wedge
\varrho_B$. \newline Furthermore  $a \varrho_B b \Rightarrow (a, b) \in \varrho_B \cup \eta_B \subseteq
\varrho_B \vee \eta_B $.
\begin{itemize}\item Case $a\in A\setminus B$, $b\in B$. \end{itemize}
In this case, it follows that $(a,b)\notin \varrho_B$ thus $(a,b)\notin \eta_B \cap \varrho_B = \eta_B \wedge \varrho_B$.
\newline Now assume that $\lambda_B(a)=\lambda_B(b)$. In this case $(a,b)\in \eta_B$ is true by definition which implies $(a,b)\in \eta_B \cup \varrho_B \subseteq \eta_B \vee \varrho_B$. \newline Otherwise: $\lambda_B(a) \neq \lambda_B(b) \Rightarrow \exists c \in B: \medskip \lambda_B(a) = \lambda_B(c)$ because $\lambda_B$ maps onto every element B. Thus $(a,c)\in \eta_B$ and $(c,b)\in \varrho_B$, this implies $(a,b)\in (\varrho_B \cup \eta_B)^T=\varrho _B\vee \eta _B$ by definition.
\begin{itemize}
\item Case $a, b \in A\setminus B$.\end{itemize}
$(a, b)\notin \varrho_B \Rightarrow (a, b)\notin \varrho_B \cap \eta_B=\varrho_B \wedge \eta_B$.
Since $\lambda _B$ maps $A$ onto $B$, thus exists such $c$ and $d$ elements of $B$ that
$\lambda_B(a)=\lambda_B(c)$ and $\lambda_B(b)=\lambda_B(d)$ holds. From $(a,c)\in \eta_B$, $(c,d)\in \varrho_B$, $(b,d)\in \eta_B$ follows
$(a ,b)\in (\varrho_B \cup \eta_B)^T =\varrho_B\vee \eta _B$.\hfill\openbox

\end{section}

\begin{section}{Boolean-type retractable automata}

\begin{defi}\label{booldef}We say that a retractable automaton $\textbf{A}=(A,X,\delta)$ is
Boolean-type if there exists a family $\{ \lambda_B \mid \bf{B} \; \textrm{is a subautomaton of} \; \bf{A} \}$ of retract homomorphism $\lambda _B$ of $\bf A$ such that, for arbitrary $\bf{B}_1$ and $\bf{B}_2$ subautomata of $\bf A$, the condition
$B_1 \subseteq B_2$ implies $Ker\lambda_{B_2} \subseteq Ker\lambda_{B_1}$.

\end{defi}

In the next, if we suppose that \textbf{A} is a Boolean-type retractable automaton and $\bf C$ is a subautomaton of $\bf A$, then $\lambda _C$ will denote the retract homomorphism of $\bf A$ onto $\bf C$ belonging to a fix family $\{ \lambda_B \mid \bf{B} \; \textrm{is a subautomaton of} \; \bf{A} \}$ of retract homomorphisms $\lambda _B$ of $\bf A$ satisfying the conditions of Definition \ref{booldef}.

In this section we shall discuss Boolean-type retractable state-finite automata without outputs.
We describe these automata using the concepts and constructions of \cite{nagy1}.

\begin{defi}We say that an automaton ${\bf A}=(A, X, \delta)$ is a direct sum of automata $\{ {\bf A}_i=(A_i,X,\delta _i)$ ($i\in I$) $\}$ (indexed with the set $I$) if
$A_i\cap A_j=\emptyset$ for every $i, j\in I$ with $i \neq j$, and moreover $A=\displaystyle\mathop{\cup}_{i\in I}A_i$.
\end{defi}

\begin{theorem}\label{7} (\cite{nagy1}) For a state-finite automaton ${\bf A}=(A,X,\delta)$ the following statements are equivalent:
\item {(i)} $\bf A$ is retractable.
\item {(ii)} $\bf A$ is the direct sum of finitely many state-finite retractable automaton, which contain kernels being isomorphic to
eachother.\hfill\openbox
\end{theorem}

\medskip
\noindent The next lemma will be used in the proof of Theorem~\ref{xxxx} several times.

\begin{lemma}\label{xxx} If $\bf D\subseteq \bf B$ are subautomaton of a Boolean-type retractable automaton $\bf A$ such that $\lambda _{B}(a)\in D$ for some $a\in A$ then $\lambda _{B}(a)=\lambda _{D}(a)$.
\end{lemma}

\noindent
{\bf Proof}. Let $c=\lambda _{B}(a)$. As $c\in D\subseteq B$, we have $\lambda _{B}(c)=c$. Thus $a$ and $c$ are in the same $Ker\lambda _{B}$-class of $A$. As every $Ker \lambda _B$-class is in a $Ker \lambda _D$-class, we have that $a$ and $c$ are in the same $\lambda _D$-class and so
$\lambda _D(a)=\lambda _D(c)$. As $c\in D$, we have $\lambda _D(c)=c$ and so $\lambda _D(a)=\lambda _D(c)=c=\lambda _B(a)$.

\begin{theorem}\label{xxxx} For a state-finite automaton ${\bf A}=(A,X,\delta)$ the following statements are equivalent:
\item {(i)} $\bf A$ is a  Boolean-type retractable automaton.
\item {(ii)} $\bf A$ is the direct sum of finitely many state-finite Boolean-type retractable automata containing kernels being isomorphic to each other.
\end{theorem}

\noindent \textbf{Proof} (i) $\mapsto$ (ii): Let $\bf A$ be a Boolean-type, retractable, state-finite automaton. Since $\bf A$ is state-finite and retractable, then by Theorem \ref{7} $\textbf{A}$ is a
direct sum of finitely many, state-finite, retractable  automata $\bf
A_i$ ($i\in I$) containing kernels being isomorphic to each other.
Let $i_0 \in I$ be an arbitrary fixed index. Let $\{ \lambda_B \mid
\textbf{B} \; \textrm{is a subautomaton of} \; \textbf{A} \} $ be a
family of retract homomorphisms such that $B_1 \subseteq B_2$ implies
$Ker\lambda_{B_2} \subseteq Ker\lambda_{B_1}$. It is clear that
$A_{i_0}$ is a subautomaton of $A$. Consider those $\lambda_C$ retract
homomorphisms which fulfils $C \subseteq A_{i_0}$, we shall denote
these with $\{ \Lambda_C \mid C \subseteq A_{i_0} \}$. Since all
\textbf{C} subautomata of \textbf{A} that has $C \subseteq A_{i_0}$
are also subautomata of $\bf{ A}_{i_0}  $, therefore the family $\{ \Lambda_C \}$
clearly fulfils the condition $Ker\Lambda_{C_2} \subseteq
Ker\Lambda_{C_1}$ for all $C_1 \subseteq C_2 \subseteq A_{i_0}$.
\medskip
\\ \noindent (ii) $\mapsto$ (i):
Assume that the automaton $\bf A$ is a direct sum of Boolean-type retractable automata $\bf A_i$ ($i\in I=\{1, 2, \dots , n\}$) whose kernels $\bf T_i$ are isomorphic to each other. Let $(\cdot )\varphi _{i, i}$ denote the identical mapping of $T_i$ ($i=1, \dots , n$). For arbitrary $i=1, \dots n-1$, let
$(\cdot )\varphi _{i, i+1}$ denote the corresponding isomorphism of $T_i$ onto $T_{i+1}$. For arbitrary $i, j\in I$ with $i<j$, let
$(\cdot )\Phi _{i, j}=\varphi _{i, i+1}\circ \cdots \circ \varphi _{j-1, j}$. For arbitrary $i, j\in I$ with $i>j$, let
$(\cdot )\Phi _{i, j}=\varphi _{i-1, i}^{-1}\circ \cdots \circ \varphi _{i, i+1}^{-1}$. It is clear that $\Phi _{i, j}$ is an isomorphism of $T_i$ onto $T_j$. Moreover, for every $i, j, k\in I$, $\Phi_{i, j}\circ \Phi _{j, k}=\Phi _{i, k}$.

Let $\bf B$ be a subautomaton of $\bf A$. Let $\cal B$ denote the set of all indexes from $1, 2, \dots , n$ which satisfy $B_i=B\cap A_i\neq \emptyset$. If $i\in {\cal B}$ then $T_i\subseteq B_i$. Let $i_B=min{\cal B}$.

We give a retract homomorphism $\Lambda _B$ of $\bf A$ onto $\bf B$. If $i\in {\cal B}$ then  let $\Lambda _B(a)=\lambda _{B_i}(a)$ for every $a\in A_i$. If $i\in I\setminus {\cal B}$ (that is, $B_i=\emptyset$) then let $\Lambda _B(a)=(\lambda _{T_i}(a))\Phi _{i, i_B}$. It is easy to see that $\Lambda _B$ is a retract homomorphism of $\bf A$ onto $\bf B$.

We show that the set $\{\Lambda _B \mid \bf B\ \hbox{is a subautomaton of}\ A\}$ satisfies the condition that, for every subautomaton
$\bf D\subseteq \bf B$, $Ker\Lambda _{B}\subseteq Ker\Lambda _{D}$. Let $\bf D\subseteq \bf B$ be arbitrary subautomata of $\bf A$. We note that $\cal D\subseteq \cal B$ and $i_B\leq i_D$. Assume
\[\Lambda _{B}(a)=\Lambda _{B}(b)\] for some $a\in A_i$ and $b\in A_j$.

\noindent
Case 1: $i\in {\cal D}$. In this case $i_D\leq i$. We have two subcases.
If $j\in {\cal B}$ then  \[\lambda _{B_i}(a)=\Lambda _B(a)=\Lambda _B(b)=\lambda _{B_i}(b)\] and so $j=i$. From this it follows that
\[\lambda _{D_i}(a)=\lambda _{D_i}(b)\] and so \[\Lambda _D(a)=\lambda _{D_i}(a)=\lambda _{D_i}(b)=\Lambda _D(b).\]
If $j\in I\setminus {\cal B}$ then
\[\lambda _{B_i}(a)=\Lambda _B(a)=\Lambda _B(b)=(\lambda _{T_j}(b))\Phi _{j, i_B}\in T_{i_B}\subseteq D_{i_B}\] and so
$i=i_B\leq i_D$. This and the above $i_D\leq i$ together imply $i=i_B=i_D$. Then
By Lemma~\ref{xxx},
\[\Lambda _D(a)=\lambda _{D_i}(a)=\lambda _{B_i}(a).\] As
\[\Lambda _D(b)=(\lambda _{T_j}(b))\Phi _{j, i_D},\]
we have \[\Lambda _D(a)=\lambda _{B_i}(a)=(\lambda _{T_j}(b))\Phi _{j, i_B}=(\lambda _{T_j}(b))\Phi _{j, i_D}\Lambda _D(b).\]

Case 2: $i\notin {\cal D}$, but $i\in {\cal B}$. If $j\in {\cal B}$ then \[\lambda _{B_i}(a)=\Lambda _B(a)=\Lambda _B(b)=\lambda _{B_i}(b)\] and so $j=i$. Then $\Lambda _D(a)=\Lambda _D(b)$ (see the first subcase of Case 1). If $j\notin {\cal B}$ then
\[\lambda _{B_i}(a)=\Lambda _B(a)=\Lambda _B(b)=(\lambda _{T_j}(b))\Phi _{j, i_B}\] and so
$i=i_B$. Thus $\lambda _{B_i}(a)\in T_i\subseteq D_i$ and so (by Lemma~\ref{xxx}) \[\lambda _{B_i}(a)=\lambda _{D_i}(a)=\lambda _{T_i}(a).\] If $i_D=i_B(=i)$ then
\[\Lambda _D(a)=\lambda _{D_i}(a)=\lambda _{B_i}(a)=(\lambda _{T_j}(b))\Phi _{j, i_B}=(\lambda _{T_j}(b))\Phi _{j, i_D}=\Lambda _D(b).\] If $i_D>i_B(=i)$ then $A_i\cap D=\emptyset$ and so
\[\Lambda _D(a)=(\lambda _{T_i}(a))\Phi _{i, i_D}\] and \[\Lambda _D(b)=(\lambda _{T_j}(b))\Phi _{j, i_D}.\] As
\[\lambda _{T_i}(a)=\lambda _{B_i}(a)=(\lambda _{T_j}(b))\Phi _{j, i_B},\] we have
\[\Lambda _D(b)=(\lambda _{T_j}(b))\Phi _{j, i_D}=(\lambda _{T_j}(b))(\Phi _{j, i_B}\circ \Phi _{i_B, i_D})=\]
\[=((\lambda _{T_j}(b))\Phi _{j, i_B})\Phi _{i_B, i_D}=(\lambda _{T_i}(a))\Phi _{i_B, i_D}=(\lambda _{T_i}(a))\Phi _{i, i_D}=\Lambda _D(a).\]

Case 3: $i\notin {\cal B}$. If $j\in {\cal B}$ then we can prove (as in the second subcases of Case 1 and Case 2) that $\Lambda _D(a)=\Lambda _D(b)$. Consider the case when $j\notin {\cal B}$. Then \[(\lambda _{T_i}(a))\Phi _{i, i_B}=\Lambda _B(a)=\Lambda _B(b)=(\lambda _{T_j}(b))\Phi _{j, i_B}.\] Hence
\[\Lambda _D(a)=(\lambda _{T_i}(a))\Phi _{i, i_D}=(\lambda _{T_i}(a))(\Phi _{i, i_B}\circ \Phi _{i_B, i_D})=
((\lambda _{T_i}(a))\Phi _{i, i_B})\Phi _{i_B, i_D}=\]
\[=((\lambda _{T_j}(b))\Phi _{j, i_B})\Phi _{i_B, i_D}=(\lambda _{T_j}(b))(\Phi _{j, i_B}\circ \Phi _{i_B, i_D})=(\lambda _{T_j}(b))\Phi _{j, i_D}=\Lambda _D(b).\]

In all cases, we have that $\Lambda _B(a)=\Lambda _B(b)$ implies $\Lambda _D(a)=\Lambda _D(b)$ for every $a, b\in A$. Consequently \[Ker\Lambda _B\subseteq Ker\Lambda _D.\] Hence $A$ is a Boolean-type retractable automaton.\hfill\openbox

\vspace{0.75cm}
\noindent By Theorem~\ref{xxxx}, we can focus our attention on a Boolean-type retractable automaton containing a kernel. In our investigation two notions will play important role. These notions are the dilation of automata and the semi-connected automata.

\begin{defi} Let ${\bf B}$ be an arbitrary subautomaton  of an automaton ${\bf A}=(A,X,\delta)$. We say that $\bf A$ is a dilation of $\bf B$ if there exists a mapping $\phi_{dil}(\cdot)$ of $A$ onto $B$ that leaves the elements of $B$ fixed, and fulfils $\delta (a,x)=\delta _B(\phi_{dil} (a),x)$ for all $a\in A$ and $x\in X$. This fact will be denoted by: (A,X,$\delta$;B,$\phi_{dil}$). (\cite{nagy3})
\end{defi}

\medskip

If $a$ is an arbitrary element of an \textbf{A} automaton, then let $R(a)$ denote the subautomaton generated by the element $a$ (the smallest subautomaton containing $a$). It is easy to see that \[R(a)= \lbrace \delta(a,x) : x \in X^* \rbrace ,\] where $X^*$ is the free monoid over $X$. Let us define the following relation: \[\mathcal{R} \coloneqq \lbrace \textrm{(a,b)} \in A \times A  : R(a)=R(b) \rbrace.\] It is evident that $\mathcal{R}$ is an equivalence relation. The $\mathcal{R}$ class containing a particular $a$ element is denoted by $R_a$. The set $R(a)\setminus R_a$ is denoted by $R[a]$. It is clear that $R[a]$  is either empty set or a subautomaton of $\textbf{A}$. $R\lbrace a \rbrace = R(a) / \rho_{R[a]}$ factor automaton is called a principal factor of \textbf{A}. If R[a] is an empty set, then consider $R\lbrace a \rbrace$ as $R(a)$. \cite{nagy1}

An \textbf{A} automaton is said to be strongly connected if, for any  $a,b \in A$ elements, there exist a word $p \in X^+$ such that $\delta(a,p)=b$; ($X^+$ is the free semigroup over $X$). Remark: for a word $p=x_1x_2\dots x_n$ and an element $a$ the transition function is defined as the following: \[\delta(a,p)=\delta \left( \dots \delta( \delta(a,x_1) ,x_2) \dots x_n \right).\] An automaton is called strongly trap connected if it contains exactly one trap and, for every $a\in A\setminus\{trap\}$ and $b\in A$, there is a word $p\in X^+$ such that $\delta (a, p)=b$.

An automaton is said to be semi-connected if its every principal factor is either strongly connected or strongly trap connected. (\cite{nagy1})

\begin{theorem}\label{csakretr} (\cite{nagy1}) A state-finite automaton without outputs is a retractable automaton if and only if it is a dilation of a semi-connected retractable automaton.\hfill\openbox
\end{theorem}

\medskip
The next theorem is the extension of Theorem~\ref{csakretr}.

\begin{theorem}\label{thm5} A state-finite automaton without outputs is a Boolean-type retractable automaton if and only if it is a dilation of a semi-connected Boolean-type retractable automaton.
\end{theorem}

\noindent{\bf Proof}. Let $\bf A$ be a Boolean-type retractable state-
finite automaton without outputs. Then, by Theorem~\ref{csakretr},
\textbf{A} is a dilation of the retractable semi-connected automaton
\textbf{C}. For a subautomaton \textbf{B} of \textbf{C}, let $\lambda_B'$ denote the restriction of $\lambda_B$ to \textbf{C}. It is easy to see that \textbf{C} is a Boolean-type retractable automaton with the family $\{ \lambda_B' \mid \textrm{\textbf{B} is a subautomaton of of \textbf{C}} \}$.
\medskip

\noindent Conversely, let the automaton
\textbf{A}=(A,X,$\delta$;B,$\phi_{dil}$)  be a  dilation of the automaton \textbf{B}=(B,X,$\delta_B$). Assume that \textbf{B} is Boolean-type retractable with the family \\ $\{\lambda_C \mid\textrm{ \textbf{C} is a subautomaton of \textbf{B}} \}$. Since all subautomata of \textbf{A} are subautomata of \textbf{B}, it is clear that, for every subautomaton \textbf{C} of \textbf{A}, $\lambda_C \circ \phi_{dil}$ is a retract homomorphism of \textbf{A} onto \textbf{C}. Moreover, \textbf{A} is a Boolean-type retractable automaton with the family $\{ \lambda_C \circ \phi \mid \textrm{\textbf{C} is a subautomaton of \textbf{A}} \}$. \hfill\openbox

\vspace{1cm}
By Theorem~\ref{thm5} and Theorem~\ref{xxxx}, we can concentrate our attention on semi-connected automata containing kernels.

\medskip

\begin{defi} Let $(T,\leq )$  be a partially ordered set, in which every two element subset has a lower bound, and every non-empty subset of $T$ having an upper bound contains a maximal element. Consider the operation on $T$ which maps a couple $(t_1,t_2) \in  T \times T $ to the (unique) greatest upper bound of the set $ \{ t_1,t_2 \} $. $T$ is a semilattice under this operation. This semilattice is called a tree. It is clear that every finite tree has a least element. (\cite{tully}) \end{defi}

If a non-trivial state-finite automaton \textbf{A} contains exactly one trap $a_0$ then $A^0$  will denote the set $A \setminus a_{0}$. If $A$ is a trivial automaton, then let $A^0 = A$. On the set $A^0 \times X $ we consider a partial (transition) function $\delta^0$ which is defined only on couples $(a,x)$ for which $\delta(a,x) \in A^0$; in this case $\delta^0(a,x)=\delta(a,x)$. We shall say that ($A^0$,X,$\delta^0$) is the partial automaton derived from the automaton \textbf{A}.

If $\textbf{A}^0$ and $\textbf{B}^0$  are partial automata, then a mapping $\phi$ of $A^0$ into $B^0$ is called a partial homomorphism of $\bf{A^0}$ into $\bf{B^0}$ if, for every $a \in A^0$ and $x \in X$, the condition $\delta_A(a,x)\in A^0$ implies $\delta_B(\phi(a),x) \in B^0$ and $\delta_B(\phi(a),x)=\phi(\delta(a,x))$.

\bigskip
\noindent \textbf{Construction} (\cite{nagy1}) Let $(T, \leq)$ be a finite tree with the least element $i_0$. Let $i\succ j$ ($i,j\in T$) denote the fact that $i\geq j$  and for all $k\in T$, the condition $i\geq k\geq j$ implies $i=k$ or $j=k$.
\noindent Let ${\bf A}_i=(A_i,X,\delta _i)$, $i\in T$ be a family of pairwise disjoint automata satisfying the following conditions:
\begin{itemize}
\item[(i)]${\bf A}_{i_0}$ is strongly connected and ${\bf A}_i$ is strongly trap connected for every $i\in T, i\neq i_0$.
\item[(ii)]Let $\phi_{i,i}$ denote the identical mapping of ${\bf A}_i$. Assume that, for every $i,j\in T, i\succ j$, there exist a homomorphism
    $\phi_{i,j}$ which maps ${\bf A}^0_i$ into ${\bf A}^0_j$ such that
\item[(iii)] for every  $i\succ j$ there exist elements $a\in A^0_i$ and $x\in X$ such that $\delta _i(a,x)\notin A^0_i$, $\delta _j(\phi _{i,j}(a),x)\in A^0_j$.
\end{itemize}

For arbitrary elements $i,j\in T$ with $i\geq j$, we define a partial homomorphism $\Phi _{i,j}$ of ${\bf A}^0_i$ into ${\bf A}^0_j$ as follows:
$\Phi_{i,i}=\phi _{i,i}$ and, if $i>j$ such that $i\succ k_1\succ \dots k_n\succ j$, then let $$\Phi _{i,j}=\phi _{k_n,j}\circ \phi _{k_{n-1},k_n} \circ \dots \circ \phi _{k_1,k_2}\circ \phi _{i,k_1}.$$ (We note that if $i\geq j\geq k$ are arbitrary elements of $T$, then $\Phi _{i,k}=\Phi _{j,k}\circ \Phi _{i,j}$.)

Let $A= \displaystyle \mathop \cup_{i\in T}A^0_i$. Define a transition function $\delta ':A\times X \mapsto A$ as follows.
If $a\in A^0_i$ and $x\in X$
then let \[\delta '(a,x)=\delta _{i'[a,x]}(\Phi _{i,i'[a,x]}(a),x),\] where
$i'[a,x]$ denotes the greatest element of the set
$\{ j\in T:\ \delta _j(\Phi _{i,j}(a),x)\in ~ A^0_j\}.$
It is clear that  ${\bf A}=(A,X,\delta ')$ is an automaton which will be denoted by $(A_i,X,\delta _i;\phi_{i,j},T)$.

\begin{theorem} \label{yyy} (\cite{nagy1}) A state-finite automaton without outputs is a semi connected retractable automaton containing a kernel if and only if it is isomorphic to an automaton $(A_i,X,\delta _i; \phi_{i,j},T)$  defined in the Construction.\hfill\openbox \end{theorem}
 \begin{remark}\label{rem1}\rm
 By the proof of Theorem 7 of \cite{nagy1} if \textbf{R} is a subautomaton of an automaton $(A_i,X,\delta _i; \phi_{i,j},T)$ constructed as above, then there is an ideal $\Gamma \subseteq T $ such that $R= \displaystyle \mathop \cup_{j \in \Gamma} A_{j}^{0}$. As T is a tree $$ \pi : i \mapsto  max\{ \gamma \in \Gamma : \gamma \leq i  \}$$ is a well defined mapping of T onto $\Gamma$ which leaves the elements of $\Gamma$ fixed. $\lambda_R$ defined by $\lambda_R(a) = \Phi_{i, \pi(i)} (a) $ ($a \in A_{i}^0$ ) is a retract homomorphism of \textbf{A} onto \text{R}. (\cite{nagy1})
This fact will be used in the proof of the next Theorem.\end{remark}

\begin{theorem} A state-finite automaton without outputs is a semi-connected Boolean-type retractable automaton containing a kernel if and only if it is isomorphic to an automaton $(A_i,X,\delta _i; \phi_{i,j},T)$  defined in the Construction. \end{theorem}

\noindent \textbf{Proof} Let $\bf A$ be a state-finite automaton without outputs which contains a kernel. Assume that $\bf A$ is also semi-connected and Boolean-type retractable. Then, by Theorem \ref{yyy}, $\bf A$ is isomorphic to an automaton \textbf{A}=$(A_i,X,\delta _i; \phi_{i,j},T)$ which is defined in the Construction.

The main part of the proof is to show that every automaton \textbf{A}=$(A_i,X,\delta _i; \phi_{i,j},T)$ constructed as above is Boolean-type retractable.
According to Theorem~\ref{yyy} the automaton \textbf{A}=$(A_i,X,\delta _i; \phi_{i,j},T)$ is retractable.
Let \textbf{B} be a subautomaton of \textbf{A}. By Remark \ref{rem1} there is an ideal $\Gamma \subseteq T$  such that $B=\displaystyle{\mathop \cup_{j \in \Gamma} }A^0_j$. Let $\pi_B : i \mapsto \{ \gamma \in \Gamma : \gamma \leq i \}$. For every $a \in A_j (j\in T)$ let $\lambda_B(a)\coloneqq \Phi_{j,\pi(j)}(a)$. Using also Remark \ref{rem1}, it is easy to see that $\lambda_B$ is a retract homomorphism of \textbf{A} onto \textbf{B}. Let $B_1$ and $B_2$ be arbitrary subautomata with $B_1\subseteq B_2$. We will show that $Ker\lambda_{B_2} \subseteq Ker\lambda_{B_1}$. Assume $\lambda_2(a)=\lambda_2(b)$ for some $a,b \in A$. According to Remark \ref{rem1}, $\lambda_{B_1}= \Phi_{ \pi_{B_2}(j),\pi_{B_1}(j)} \circ \Phi_{j, \pi_{B_2}(j)}$. Thus

$$ \lambda_{B_1}(a) = (\Phi_{ \pi_{B_2}(i),\pi_{B_1}(i)} \circ \Phi_{j, \pi_{B_2}(i)}) (a) = (\Phi_{ \pi_{B_2}(i),\pi_{B_1}(i)} \circ \lambda_{B_2}) (a) =$$ $$ =  (\Phi_{ \pi_{B_2}(i),\pi_{B_1}(i)} \circ \lambda_{B_2}) (b) =(\Phi_{ \pi_{B_2}(i),\pi_{B_1}(i)} \circ \Phi_{j, \pi_{B_2}(i)}) (b)= \lambda_{B_1}(b).$$ Consequently $Ker \lambda_{B_2} \subseteq Ker \lambda_{B_1} $. Hence \textbf{A}=$(A_i,X,\delta _i; \phi_{i,j},T)$ is a Boolean-type retractable automaton with the family $\{\lambda_B \mid \textrm{\textbf{B} is a subautomaton of \textbf{A}}\}$. \hfill\openbox


\end{section}

\end{document}